# Heterogeneous Multi-core Array-based DNN Accelerator

Mohammad Ali Maleki[1], Mehdi Kamal[2], Ali Afzali-Kusha[1, 3]
[1]School of Electrical and Computer Engineering, College of Engineering, University of Tehran
[2]Electrical Engineering Department, University of Southern California
[3]Computer Science, Institute for Research in Fundamental Sciences
m.a.maleki@ut.ac.ir, mehdi.kamal@usc.edu, afzali@ut.ac.ir

***Abstract*—in this article, we investigate the impact of architectural parameters of array-based DNN accelerators on accelerator's energy consumption and performance in a wide variety of network topologies. For this purpose, we have developed a tool that simulates the execution of neural networks on array-based accelerators and has the capability of testing different configurations for the estimation of energy consumption and processing latency. Based on our analysis of the behavior of benchmark networks under different architectural parameters, we offer a few recommendations for having an efficient yet high-performance accelerator design. Next, we propose a heterogeneous multi-core chip scheme for deep neural network execution. The evaluations of a selective small search space indicate that the execution of neural networks on their near-optimal core configuration can save up to 36% and 67% of energy consumption and energy-delay product respectively. Also, we suggest an algorithm to distribute the processing of network's layers across multiple cores of the same type in order to speed up the computations through model parallelism. Evaluations on different networks and with the different number of cores verify the effectiveness of the proposed algorithm in speeding up the processing to near-optimal values.**

## I. Introduction

DEEP neural networks (DNN) have become one of the most effective, yet reliable algorithms in a broad range of applications. Self-driving cars [1], [2], drug discovery [3], [4], DNA and human genome study and their role in diseases[5], [6], healthcare and medicine [7], [8], robotics [9], computer games [10], [11], speech recognition and natural language processing [12], [13] are a few examples that DNNs have been deployed successfully with comparable or even better results than other techniques. Convolutional neural networks (CNNs) are a subset of DNNs that play a major role in image and speech recognition with tremendous results on large-scale datasets [14]–[20].

The human-level accuracy of deep CNNs comes at a very high energy cost related to the massive number of computations and memory accesses during the processing of a single input sample [21], [22]. This high energy usage is prohibitive for the deployment of deep networks in embedded systems with limited power budget. In addition, the energy consumption of large-scale neural models in enterprise implementations (*e.g.*, data centers), has a direct impact on their operating expenses mainly due to the more computation loads on servers and consequently, higher energy consumption of their cooling systems [21].

Solutions to reduce energy consumption in neural networks would span from algorithmic level down to microarchitecture, hardware, and even devices. From algorithm level perspective, many efforts have been made toward lighter network topologies while maintaining final accuracy at the desired level in recent years. Along with these modifications in network topologies, there are a number of other methods that affect the network layers but are usually applied after the topology of the network has been determined. Pruning the weights or in a more radical manner, pruning a whole kernel in network layers while keeping the accuracy untouched or above a predefined desired level is one of these methods that can diminish both computational and memory access costs resulting in overall lower energy consumption [22]–[24].

There is a variety of other methods that applies at the runtime in favor of decreasing the energy consumption or latency of the processing. Encoding and compression [25]–[27], using approximate computing [28], [29], utilizing quantized or truncated values in the processing of the networks [30]–[33], changing dataflow and re-ordering the computations to reduce the memory access can be classified in this category of works [22], [34].

In spite of the fact that the aforementioned methods have an impact on energy reduction, as long as the network is deployed on general-purpose units, the corresponding energy cost of the system remains above the acceptable threshold for many applications. Domain-specific architecture (DSA) is a response to the need for more efficient yet high-performance processing units in the post-Dennard law era. There are many neural network accelerator designs that each one tries to optimize an objective such as energy, latency, throughput, or a combination of these factors in processing the neural networks either targeting the FPGA as the implementation platform or using an ASIC chip design [35]–[38].

Despite numerous different designs for neural network accelerators, the array-based accelerator, in which a number of processing elements are placed in a two-dimensional arrangement, is the dominant scheme. Even though each work has its pros and cons compared to others, the impact of each architectural element on the overall performance of the system is not studied comprehensively. To address this issue, in this paper we explore the impact of array-based accelerator's architectural elements impact on the final efficiency and performance of different neural networks processing and based on this analytical study, we propose a high-level scheme for multi-core array-based accelerators that can operate at near-optimal points for diverse categories of neural networks.

In summary, the contribution of the paper is as follows:

- Developing a high-level tool for the first-order estimation of the energy and latency of different network topologies executed on the array-based accelerators (Section II. the developed tool).
- Exploring the impact and study of accelerator architectural changes on the final energy consumption and performance (equivalently the latency) of the accelerator during the processing of the networks (Section III. the effect of accelerator's architectural elements variations on the efficiency and latency of processing).
- Proposing a high-level heterogeneous multi-core processing chip scheme for efficient processing of neural networks of different categories based on the mentioned study (Section IV. heterogeneous multi-core chip for efficient processing of neural networks).
- Proposing a simple algorithm to distribute the processing of network layers among the multiple cores of one type (a.k.a. homogeneous cores) to speed up the computations through model parallelism (Section IV.B. model parallelism on homogeneous cores).

## II. THE DEVELOPED TOOL

From the highest point of view, the exploration space of the path to designing an efficient accelerator, depending on the metrics of concern, for a wide range of neural network topologies[1], depends on I) the neural networks topology, II) the microarchitecture of the accelerator and III) the mapping of the network onto the accelerator also known as dataflow (Fig. *1*). Reliable tools have been developed for finding a proper topology of neural network for a particular application considering the characteristics defined by the user, such as the minimum required accuracy or the size limit of the network [39], [40]. Moreover, pre-trained networks on a specific dataset are available in public repositories and are fine-tuned on the desirable target dataset.

After deciding on the neural network topology, the network layers should be mapped to the accelerator for processing. The mapping has correlation with hardware architecture; that is, the hardware platform must support the dataflow by which the neural network will be processed. Regardless of the specific details that each design may have, Fig. 2 shows a simplified structure of the array-based accelerator used as the processing core in our developed tool.

[1] Since the paper discuss the accelerator's hardware architecture, we preferred to use neural network ***topology*** instead of neural network ***architecture*** to avoid confusion.

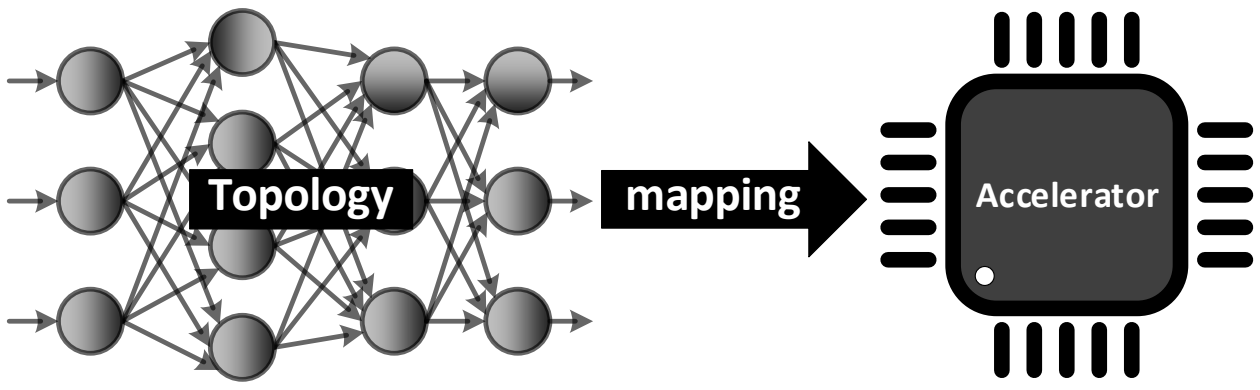

Fig. 1. The state space for designing an efficient neural network.

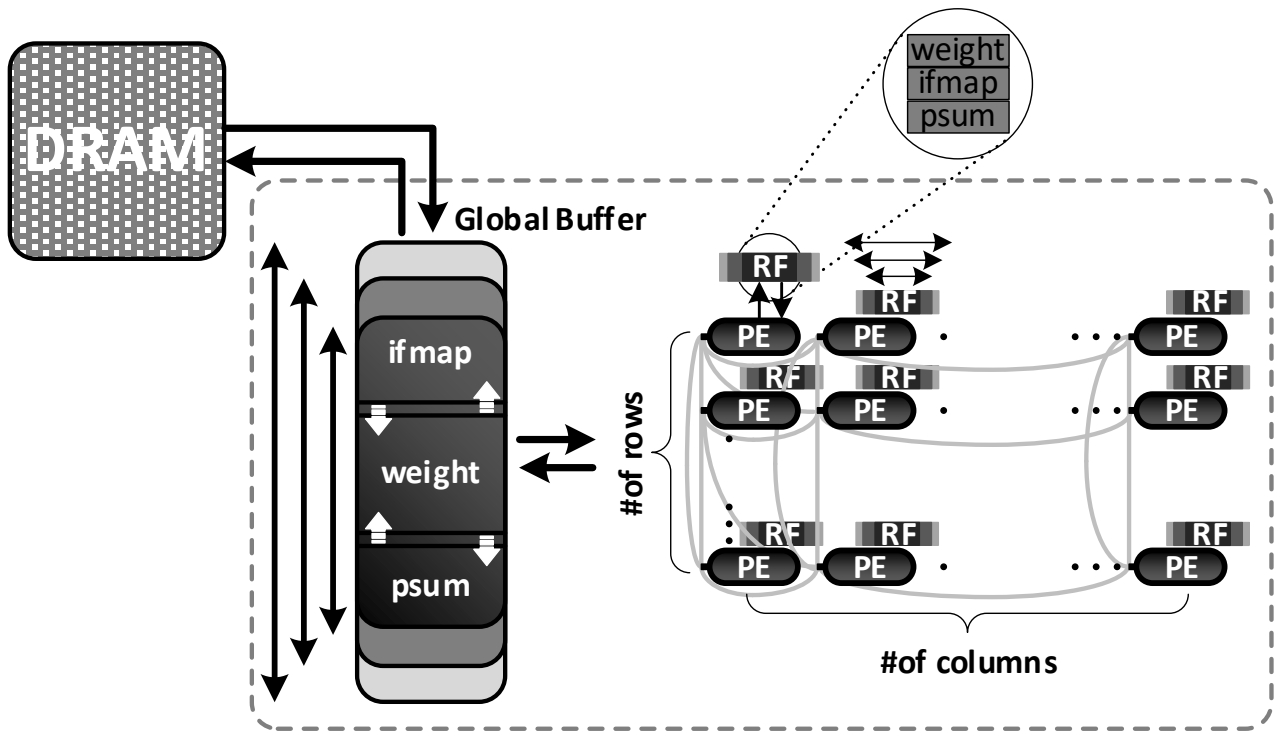

Fig. 2. The overall structure of an array-based accelerator.

This configuration consists of processing elements (indicated by PE) and memory elements (RF, Global Buffer, and off-chip DRAM) arranged in a specific hierarchy. Each processing element has its own local register file (also known as *scratch pad*) for the storage of the data being processed in order to keep it as close as possible. A network-on-chip can be utilized for data transfer between the PEs or they can be connected to each other by simpler methods like mesh grid. Depending on the requirements and design constraints, data can be transferred in various forms such as multicast, unicast, and broadcast. Global buffer communicates with processing elements via network-on-a-chip (or mesh) and provides their needed data. The array's processed data can be stored in the on-chip global buffer as an intermediate storage element unless it needs to be stored on off-chip DRAM due to insufficient storage space in the buffer or for writing the final results of a processed layer. Off-chip DRAM stores all the weights and inputs of a specific layer that are going to be processed, as well as the processed outputs of these layers, which will be used as the input of the next layer. The energy cost of the memory hierarchy from register files to DRAM is incremental. Approximately, the DRAM energy consumption for single access is about several tens of times that of local RFs whereas the global buffer consumes about 5 to 10 times that of the local file register energy consumption.

A convolution layer depicted in Algorithm I is composed of six nested for-loops that convolve the inputs with the filters (also known as kernels). Dataflow determines the order of execution and processing of these nested loops, taking into account the potential and capacity of the hardware that executes it. Since the order of execution determines how data is re-used in different levels of the memory hierarchy, therefore it has a direct effect on the amount of energy consumption and latency of the processing. As it has been discussed in [41], row-stationary dataflow tries to maximize the use of lower-level memory elements in the memory hierarchy (equivalently, with lower energy consumption and simultaneously faster), resulting in better energy efficiency compared to other well-known dataflows. Thus, we utilized this dataflow in the developed tool.

**ALGORITHM I** THE CONVOLUTION OPERATION**

**for** $m=1:M$ **do**
  **for** $c=1:C$ **do**
    **for** $i_x=1:stride:(I_x - K_x + 2\times pad)/stride + 1$ **do**
      **for** $i_y=1:stride:(I_y - K_y + 2\times pad)/stride + 1$ **do**
        **for** $k_x=1:K_x$ **do**
          **for** $k_y=1:K_y$ **do**
            $O[m][i_x][i_y] += filter[m][c][k_x][k_y]\times I[c][i_x+k_x][i_y+k_y]$

**$M$ refers to the number of the input feature map, and $C$ refers to the number of channels in both the inputs and filters. Filters are in $K_x\times K_y$ size.

*A. Processing core's energy and latency estimation*

Processing core's energy and latency can be divided into two distinct categories: the energy and latency of the computational part (i.e. The processing array's energy and latency) and the energy and latency of the memory elements in the memory hierarchy. In other words, by having the energy consumption and latency of each memory element and counting the number of accesses (read/write) to it which is governed by dataflow, the energy consumption and latency of the memory hierarchy can be estimated. In the same manner, by having the energy and latency of each processing element, the energy consumption and latency of the processing array can be estimated by considering the processing's array utilization (i.e. number of active PEs in each processing pass) and data transfer between PEs, which is controlled by a specific protocol. The following subsections describe these two calculations in detail.

*A.1. Processing core's energy estimation*

Energy is cumulative, which means that the total energy consumed is the sum of the energy of each data movement in the memory hierarchy (reading or writing data at each memory level) and each execution by each processing element. Each data movement at each level consists of a read operation from one level and a write operation to another level of memory. Fig. 3 shows an example of a three-level memory hierarchy and possible data movement at different levels. The set of equations in (1) shows the energy consumption of possible data movement shown in Fig. 3.

$$
\begin{aligned}
&E_{DRAM\ to\ GB} = E_{read\ from\ DRAM} + E_{write\ to\ GB}\\
&E_{GB\ to\ RF} = E_{read\ from\ GB} + E_{write\ to\ RF}\\
&E_{RF\ to\ PE} = E_{read\ from\ RF} + E_{PE}\\
&E_{RF\ to\ GB} = E_{read\ from\ RF} + E_{write\ to\ GB}\\
&E_{GB\ to\ DRAM} = E_{read\ from\ GB} + E_{write\ to\ DRAM}
\end{aligned}
\qquad (1)
$$

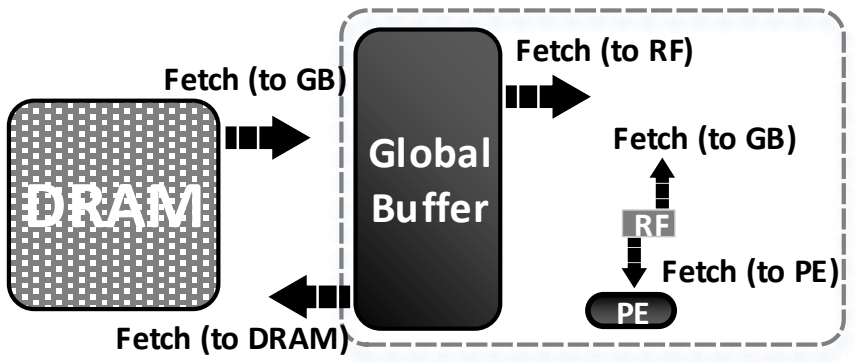

Fig. 3. Possible data movement throughout a three-level memory hierarchy.

The energy consumption of each level of memory depends on the size of the memory, the memory mode (read or write), the memory interface, the feature size in which the memory is fabricated, and the memory architecture parameters (e.g. bus width, block size, number of read/write ports). All these parameters will be defined in the tools specific for memory modeling (e.g. CACTI). The main operation of the processing element is multiply-accumulate (a.k.a. MAC). The energy of each MAC operation depends on a number of parameters such as the adding and multiplying algorithm, the bit width of operation, and the library which the design is being synthesized with.

Our developed tool keeps track of every single data movement and processing done in processing elements during the network layers' processing and computes the corresponding energy consumption of that layer.

*A.2. Processing core's latency estimation*

Similar to energy estimation, the latency calculation is also divided into two parts. The time that is spent on delivering data to the lower-level memory elements or writing back the result to higher-level memory elements, plus the time the processing array will spend on data processing to produce a valid output. In other words, as presented in (1) for the energy consumption, core's latency is the result of time spent on reading the data from DRAM and writing it to the global buffer, reading the data from the global buffer and writing it to the register files of the processing elements (this delay depends on how the data is delivered by NoC/mesh), reading the data from the register files and processing it by the array (including the time it takes to write the result back to the register file), reading from the register files and writing back to the global buffer, and finally reading from the global buffer and writing back the result on the off-chip DRAM.

It should be noted that, latency of a layer is defined as the processing of all data in DRAM and writing back generated valid outputs to DRAM. Unlike energy, latency is not cumulative. In fact, the latency is highly dependent on the dataflow controller, which determines the start time of the data fetch and the start time of processing by the processing elements in the array, as well as the network-on-chip or mesh grid architecture between the processing elements.

For a better understanding of these factors and how the tool considers them, Fig. 4 shows two hypothetical scenarios. We assume that all required data is already stored in the global buffer and two convolution operations are performed in two arrays with the row-stationary dataflow. It is also assumed that NoC is designed in such a way that each processing element can fetch its own data whenever the data is on the shared bus.

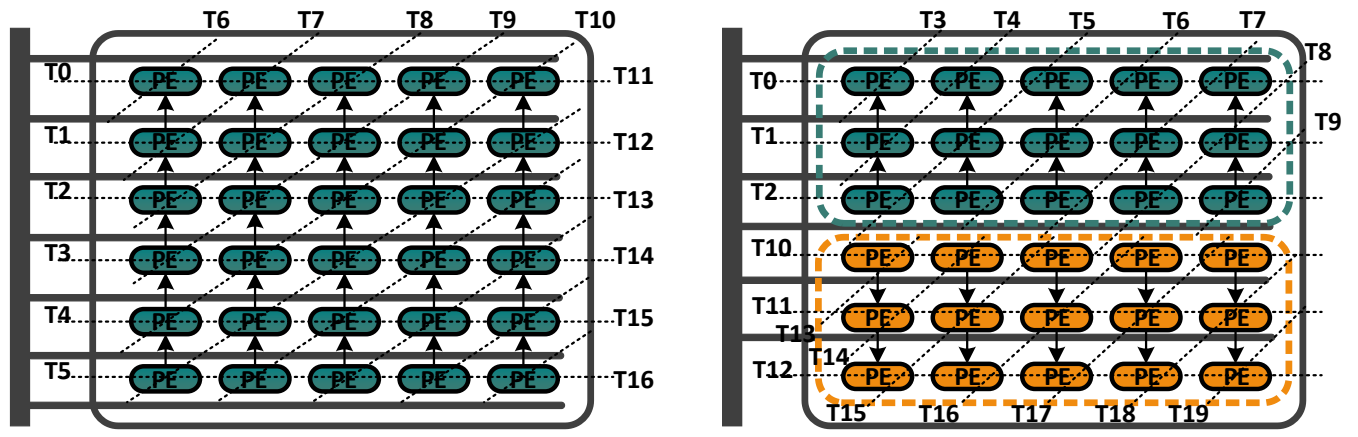

Fig. 4. The time order in which data is being delivered to processing elements. (a) The array is dedicated to one convolution processing (b) Two distinct convolutions is being processed by the array simultaneously.

In the row-stationary dataflow, all processing elements in a row receive the same row of filters for processing, while the input feature map rows are diagonally distributed to the processing elements in the array. The time labels written in the two examples in Fig. 4 show the order in which the data is placed on the shared bus by NoC. In Fig. 4(b), the array is divided to perform two convolution operations in parallel (two operations are separated with two different colors) whereas in Fig. 4(a) the entire array is dedicated to a larger convolution processing. As can be seen, the data arrival time is different in these two scenarios. Assuming that processing does not start unless the last processing element in the sub-array responsible for convolution calculations receives its data, a valid output at T10 and T20 for the upper and lower convolutions in Fig. 4(b) is produced respectively. The convolution in Fig. 4(a) is completed at time T17.

### *B. Tool's input and output*

The developed tool needs some inputs that should be provided by the user for the simulation of the neural network on the accelerator. Subsequently, the tool returns some outputs showing the performance of the accelerator. This subsection enumerates these inputs/outputs.

#### *B.1. Tool's inputs*

- Obviously, the first required input is the neural network topology. The tool accepts the neural network in a specific predefined format. In this format, the neural network structure is divided into two separate parts, one consists of convolutional layers, and the other contains fully-connected layers. The layers of each part must be defined as a separate structure. The type of layer can be one of the following, 1) input, 2) convolution, 3) subsampling (a.k.a. pooling), 4) depth-convolution, 5) point-wise convolution. Having the layer's type specified, corresponding parameters of that layer should be defined. The number of channels, number of filters, kernel size, convolution stride, padding size, and pooling stride are some examples of these parameters. Since defining a deep network based on the specified format may be a bit time-consuming, the tool is accompanied by the most commonly used neural network topologies (including 18 well-known networks). In addition, the tool is equipped with a converter that can convert the networks defined in Keras API format to their specified format automatically.
- Global buffer size and the share of each data type in its total capacity, i.e. input feature maps (ifmaps), partial sums (psum), and weights (or filter/kernel). Read/write energy consumption per access, read/write time per access.
- Array size; That is, the number of rows and columns and the energy/latency of each processing element (PE).
- Network-on-chip (or mesh) configuration and how it delivers data to the PEs.
- Registry files (scratch pads) size and the share of each aforementioned data type in its total capacity and the read/write energy consumption and read/write time per access of register file.
- DRAM, global buffer, and register file interface
- Storage and computation bit-width

It should be noted that we used CACTI to calculate the energy consumption and latency per access of memory elements. For the convenience of the users, the tool is accompanied by memory elements in various sizes with their latency and energy consumption values. Synopsis Design Compiler synthesis tool was used for obtaining the energy and latency of the MAC unit in processing elements.

#### *B.2. Tool's outputs*

By providing the inputs required by the tool, the target neural network is analyzed layer by layer and the following outputs are extracted by the tool and are provided to the user:

- The number of read and write accesses of each data type for each memory level
- Processing latency of each layer and also the amount of time spent on processing and spent on memory hierarchy.
- Energy consumption of each layer and also energy consumption of memory elements and energy consumption of processing array
- Throughput and utilization of the accelerator

## III. THE EFFECT OF ACCELERATOR'S ARCHITECTURAL ELEMENTS VARIATIONS ON THE EFFICIENCY AND LATENCY OF PROCESSING

In the introduction section, we articulated the design shift toward neural network accelerator design for more efficient processing in recent years, emphasizing the array-based accelerators as the dominant design scheme mainly due to the nature of operations (e.g. convolutions) in neural networks. This section investigates the effect of changes in the accelerator's architectural elements on the efficiency and performance of different neural networks executed on them. In particular, we will examine global buffer size changes as well as changes in the processing array size on the energy consumption and latency of neural network processing.

The global buffer acts as an intermediate memory with less power consumption than the main off-chip memory. This lower-level memory stores the weights, input feature maps, and partial sums produced by the array during the processing that may be re-used repeatedly by the array during completing the processing of a single layer convolutions.

In most of the hardware accelerator designs, a particular part is assigned for each data type (i.e. weights, input feature map, and partial sums). In view of the fact that the overall storage size necessary to store weights is relatively small compared to the storage size needed to store input and output feature maps, we further assume that the space provided for weights is constant and large enough to store all the weights required for ongoing processing in the array. In this section, we will use $GB_{psum}$ and $GB_{ifmap}$ to refer to the separate storage part of partial sums and the input feature maps, respectively, and examine the effect of their size changes alongside the arrays of different sizes on the overall performance and efficiency of the accelerator.

Fig. 5 depicts the accelerator energy consumption (for VGG16) when sweeping $GB_{psum}$ size for different array sizes and at $GB_{ifmap}$ = 216KB. As can be seen in the figure, for each given array size there is a point at which the accelerator energy consumption is minimum. The reason for this behavior is related to the amount of partial sum generated by the array, followed by the requirement to store them in the global buffer for re-use which is dictated by dataflow. In fact, if the $GB_{psum}$ capacity is not sufficient for the whole partial sums generated by the array during one pass of processing, the data must be stored in off-chip DRAM, which will incur additional costs for both writing and re-reading of the data in the next processing cycle.

For example, for the [4,4] array size, the minimum energy consumption occurs for $GB_{psum}$ = 27K. In fact, this size had been large enough to avoid partial sums from being written on off-chip memory. However, despite the lower energy consumption of 13KB buffer size, the energy consumed by the accelerator is greater than when a larger memory size with higher energy consumption (e.g. $GB_{psum}$ = 27KB ) is used. In other words, $GB_{psum}$=13KB is not sufficient to prevent data from being written on off-chip memory. Apparently, the increase in energy consumption per $GB_{psum}$ larger than 27KB (in the case of [4,4] array size) is due to the increased read/write energy of larger memories. This behavior indicates that choosing an unnecessarily larger memory size will impose additional energy costs on the system.

The same analysis can be applied to arrays of other sizes. As another example, in an array of size [16,16] we see a sudden decrease in energy consumption at two points numbered with ❶ and ❷ in Fig. 5. When the $GB_{psum}$ size increases from 27KB to 54KB, many layers do not need off-chip memory to store their partial sums generated during layer processing, resulting in lower overall energy consumption. As the amount of memory increases, the next point $GB_{psum}$ = 108KB does not show any decrease in the amount of energy consumption compared to the previous point of $GB_{psum}$ = 54KB. This means that even at $GB_{psum}$ = 108KB, there are still layers that need access to the off-chip memory for their partial sums storage. By reaching to 216KB, we see a reduction in energy consumption once again. In fact, at this point, a number of layers that needed more storage capacity on the global buffer to store their partial sums could now store their data on the buffer without having to access off-chip memory. Numerically, we see a 25% and 30% reduction in energy consumption at 54KB and 216KB points compared to the initial point of 13KB, respectively.

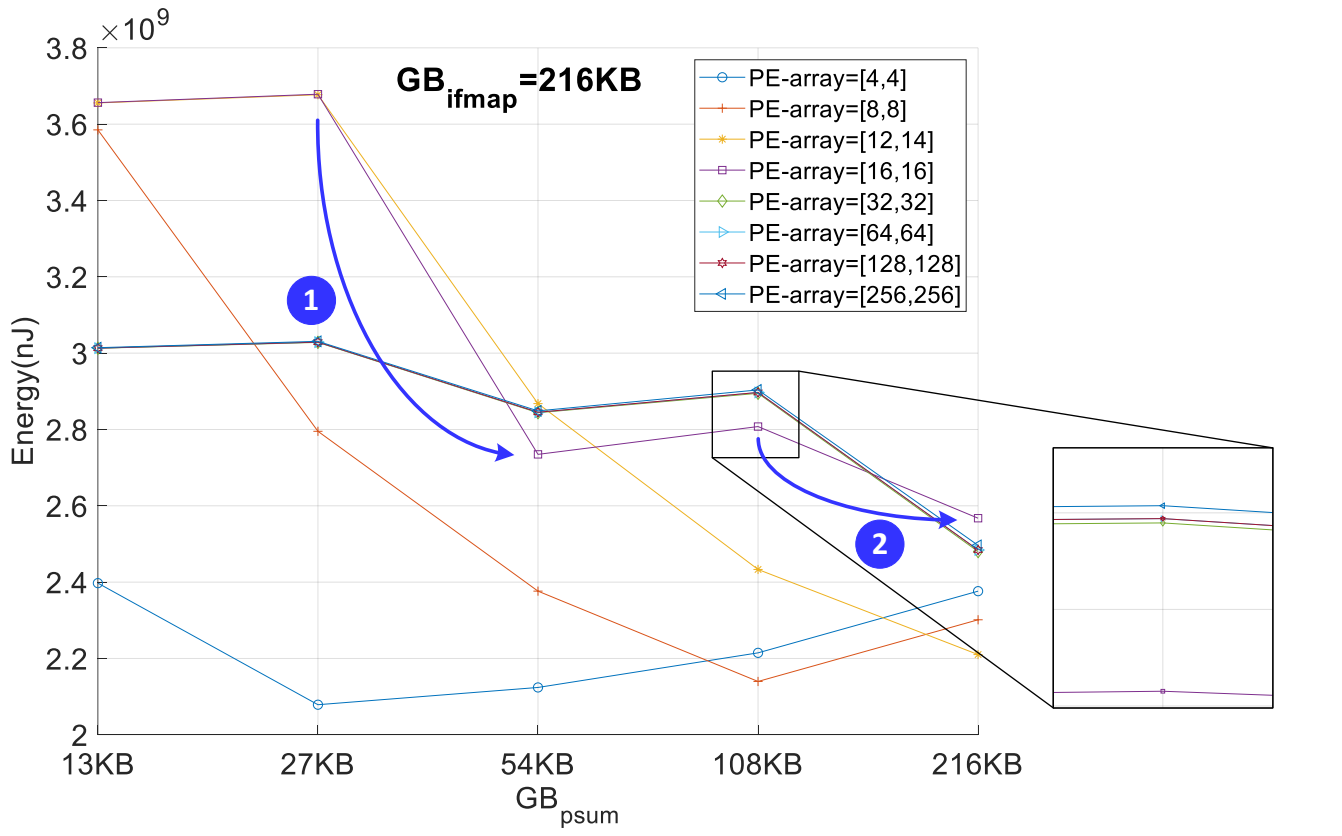

Fig. 5. Accelerator's energy consumption changes in a constant $GB_{ifmap}$ when sweeping the $GB_{psum}$ size for different array sizes.

***Observation 1:***

In a given specific dataflow, in a constant input feature map global buffer (i.e. $GB_{ifmap}$) and fixed array size, the system energy consumption is a function of partial sum buffer storage capacity ($GB_{psum}$). In other words, by considering the size of the array, it is possible to determine the partial sum buffer size so that the accelerator's energy consumption is as close as to minimum value possible.

Fig. *6* shows the accelerator's energy consumption for different $GB_{ifmap}$ values at a constant value of $GB_{psum}$ (=13KB) in different array sizes for VGG16 network. As can be seen in the figure, the energy consumption in a number of arrays is strictly monotone increasing (for example in an array size of [4,4]). This trait results from the larger buffer size's increased energy consumption. In specific points in some array sizes, a breakpoint can be seen where the system energy is decreased (a sample point is shown with ❶ in the figure). The cause of this behavior can be traced to the combined effects of $GB_{ifmap}$ capacity, array size, and dataflow. In the row-stationary dataflow implemented in the developed tool, the processing capacity of the physical array is first calculated. Processing capacity refers to the number of rows (or channels) of the input image that can be loaded to the array for processing at the same time. When the array size and the input image are in such proportion that more than one channel of the input image can be processed simultaneously, the required channels must be written from the off-chip DRAM to the on-chip global buffer and then loaded to the array for processing.

In this case, the two or more channels of the input image are processed simultaneously and the partial sums generated from each of the processed channels are added together in the array. The final result is transferred to the global buffer and finally to the off-chip DRAM. Now in this scenario, if the $GB_{ifmap}$ capacity is not sufficient to accommodate all the channels the array needs for processing, it will obviously slow down the processing in addition to the extra energy required to write the result of the processed channels back to the buffer and re-read

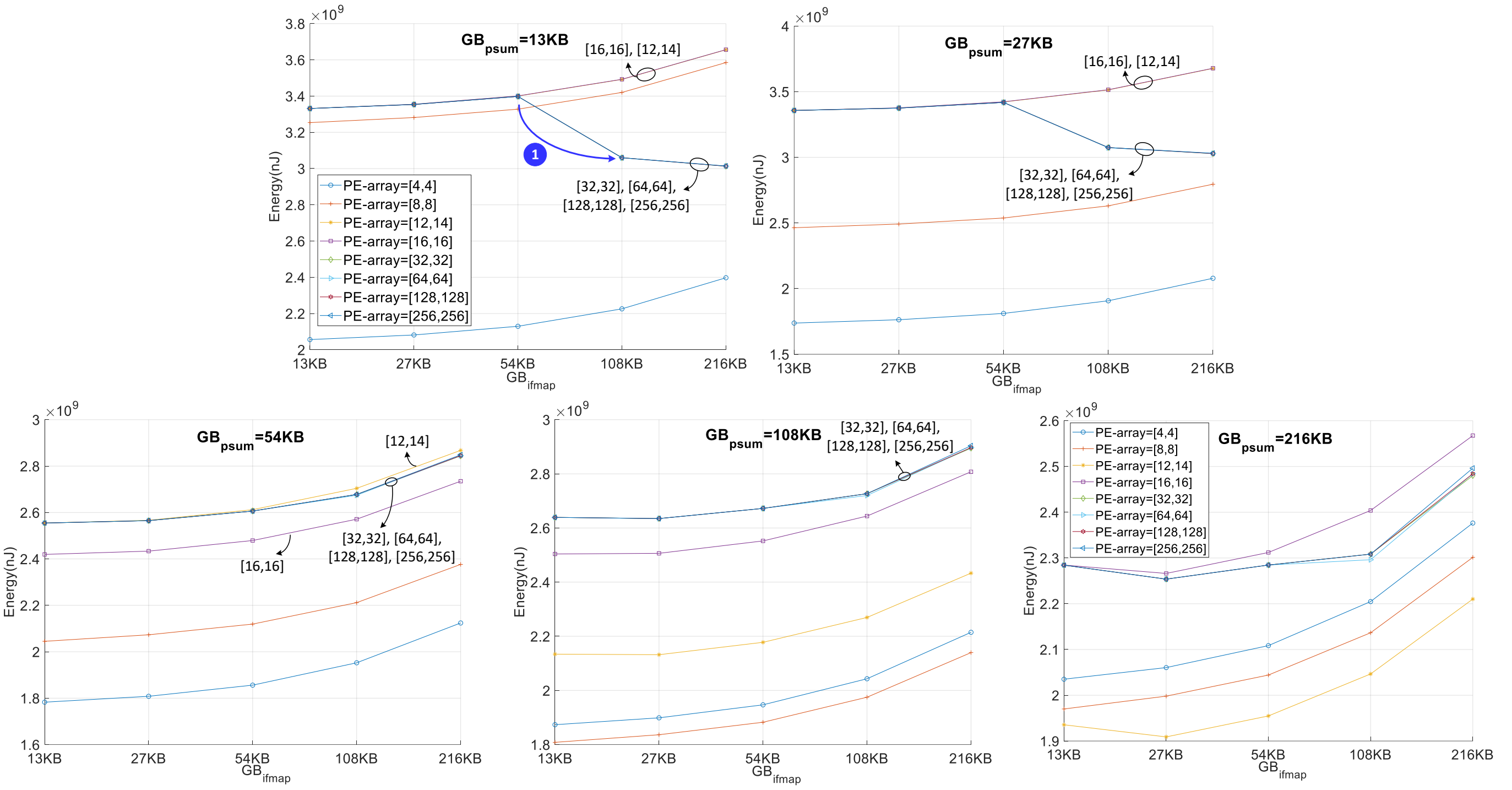

Fig. 6. Accelerator's energy consumption in a constant $GB_{psum}$ when sweeping $GB_{ifmap}$ size for different array sizes for VGG16 network.

it to add it to those that were just processed by the array.

***Observation 2:***

In a given specific dataflow, in constant $GB_{psum}$ storage space, as the processing capacity increases, the $GB_{ifmap}$ (input feature map storage space) must be large enough to provide the data needed by the array to maximize the utilization of the array. Reduced $GB_{ifmap}$ storage space, in addition to reducing array utilization, will increase the total energy consumption resulting from writing/re-reading the processing result to/from the $GB_{psum}$ storage space.

It should be noted that to increase the utilization of the array or reduce the energy consumption of writing and re-reading the processing result into the buffer, processing can be started (this is the controller's job) once all the array's required data has been moved into it. This will obviously delay the processing due to the time spent on data delivery.

To illustrate the amount of possible energy variation from the minimum point caused by changes in the size of the $GB_{ifmap}$ and $GB_{psum}$ storage spaces when one is fixed and the other is variable, we define the following parameters: the mean distance from the minimum point and the distance between the maximum and minimum point denoted by $\mu_{min}^{p}$ and, $\delta_{max}^{min}$ respectively.

$$\mu_{min}^{p} = \frac{\left(\frac{e_p - e_{min}}{e_{min}}\right) \times 100}{n-1} \tag{2}$$

$$\delta_{min}^{max} = \frac{e_{max} - e_{min}}{e_{min}} \times 100 \tag{3}$$

For sweeping the $GB_{ifmap}$ and $GB_{psum}$ variables, five values, namely $\{13KB, 27KB, 54KB, 108KB, 216\ KB\}$ are considered. It means we will have 25 search points for each fixed size of the array. In the above equations $e_{min}$ ($e_{max}$) is the amount of energy at a point where the accelerator's overall energy is at its minimum (maximum). $e_p$ is the energy of any point in the search space whose $GB_{psum}$ or $GB_{ifmap}$ size is equal to the $e_{min}$ ones. Table 1 and Table 2, respectively, show the values of these parameters when $GB_{psum}$ and $GB_{ifmap}$ are constant. The figures in these two tables illustrate that even in a more restricted search space where array size and one of the $GB_{ifmap}$ or $GB_{psum}$ are constant, a variation in the other part of the global buffer results in a significant amount of energy consumption change.

Table 3 shows the difference between the points with the highest and lowest energy consumption in the whole search space of $GB_{ifmap}$ and $GB_{psum}$ values, demoted by $\Delta_{min}^{max}$ (the search space contains 25 points of all possible combinations of values listed above).

Table 1. Accelerator's energy change in different array sizes expressed by $\mu_{min}^{p}$ and $\delta_{min}^{max}$ when $GB_{psum}$ is constant but $GB_{ifmap}$ is swept between predefined values (please refer to text)

| PE-array / Network name | [12,14] $\mu_{min}^{p}, \delta_{min}^{max}$ | [16,16] $\mu_{min}^{p}, \delta_{min}^{max}$ | [32,32] $\mu_{min}^{p}, \delta_{min}^{max}$ | [64,64] $\mu_{min}^{p}, \delta_{min}^{max}$ | [128,128] $\mu_{min}^{p}, \delta_{min}^{max}$ | [256,256] $\mu_{min}^{p}, \delta_{min}^{max}$ |
|---|---|---|---|---|---|---|
| **AlexNet** | 5.9%, 13.76% | 5.86%, 12.79% | 5.61%, 11.04% | 5.52%, 10.77% | 5.48%, 10.77% | 5.45%, 10.62% |
| **DenseNet121** | 6.46%, 17.02% | 5.85%, 15.23% | 5.82%, 14.94% | 5.93%, 14.99% | 5.99%, 15.18% | 5.98%, 15.14% |
| **DenseNet169** | 6.83%, 17.14% | 5.93%, 15.24% | 5.58%, 14.41% | 5.71%, 14.86% | 5.72%, 14.87% | 5.72%, 14.88% |
| **GoogleNet** | 3.13%, 8.18% | 2.91%, 7.6% | 2.46%, 5.96% | 2.48%, 5.95% | 2.48%, 5.95% | 2.48%, 5.95% |
| **InceptionResNetV2** | 7.28%, 17.79% | 6.73%, 16.97% | 6.54%, 16.19% | 6.49%, 16.05% | 6.49%, 16.04% | 6.51%, 16.06% |
| **InceptionV3** | 7.08%, 17.24% | 6.84%, 15.73% | 6.61%, 14.69% | 7.13%, 14.95% | 5.04%, 12.56% | 5.05%, 12.42% |
| **MobileNet** | 6.48%, 16.3% | 5.33%, 13.79% | 6.95%, 15.49% | 6.25%, 13.13% | 7.72%, 17.73% | 7.87%, 17.71% |
| **MobileNetV2** | 5.77%, 14.48% | 5.1%, 13.14% | 5.46%, 13.64% | 5.27%, 13.06% | 5.53%, 13.86% | 5.58%, 13.79% |
| **NASNetLarge** | 6.65%, 16.73% | 6.44%, 16.17% | 6.53%, 16.34% | 6.54%, 16.34% | 6.44%, 16.21% | 6.45%, 16.23% |
| **NASNetMobile** | 5.9%, 13.71% | 5.12%, 11.72% | 5.21%, 11.76% | 5.21%, 11.77% | 5.15%, 11.68% | 5.15%, 11.66% |
| **ResNet50** | 7.76%, 19.28% | 6.99%, 17.51% | 6.88%, 17.34% | 7.3%, 17.3% | 7.55%, 17.75% | 7.66%, 17.76% |
| **ResNet152** | 7.88%, 18.58% | 7.25%, 16.87% | 7.23%, 16.49% | 7.47%, 17% | 7.44%, 17% | 7.45%, 17.02% |
| **xception** | 3.25%, 7.1% | 3.68%, 7.73% | 3.51%, 6.65% | 3.51%, 6.65% | 3.59%, 6.65% | 3.66%, 6.93% |
| **VGG16** | 6.69%, 15.78% | 5.54%, 13.29% | 3.7%, 10.04% | 3.7%, 10.22% | 3.8%, 10.22% | 3.97%, 10.76% |
| **VGG19** | 6.03%, 13.13% | 5.05%, 10.96% | 5.8%, 12.75% | 6.4%, 13.4% | 5.8%, 12.75% | 5.94%, 12.75% |
| **DenseNet201** | 7.06%, 17.78% | 5.9%, 15.66% | 5.67%, 15% | 5.9%, 15.84% | 5.9%, 15.85% | 5.9%, 15.85% |

Table 2. Accelerator's energy change in different array sizes expressed by $\mu_{min}^{p}$ and $\delta_{min}^{max}$ when $GB_{ifmap}$ is constant but $GB_{psum}$ is swept between predefined values (please refer to text)

| PE-array / Network name | [12,14] $\mu_{min}^{p}, \delta_{min}^{max}$ | [16,16] $\mu_{min}^{p}, \delta_{min}^{max}$ | [32,32] $\mu_{min}^{p}, \delta_{min}^{max}$ | [64,64] $\mu_{min}^{p}, \delta_{min}^{max}$ | [128,128] $\mu_{min}^{p}, \delta_{min}^{max}$ | [256,256] $\mu_{min}^{p}, \delta_{min}^{max}$ |
|---|---|---|---|---|---|---|
| **AlexNet** | 23.3%, 51.35% | 25.1%, 45.97% | 28.44%, 53.62% | 28.44%, 53.62% | 28.44%, 53.62% | 28.44%, 53.62% |
| **DenseNet121** | 39.5%, 73.26% | 38.16%, 71.09% | 38.98%, 72.11% | 40%, 72.06% | 40%, 72.06% | 40%, 72.06% |
| **DenseNet169** | 35.4%, 74.64% | 33.89%, 70.45% | 34.48%,70.94% | 34.48%, 70.92% | 34.48%, 7.94% | 34.48%, 7.94% |
| **GoogleNet** | 71.39%, 112% | 72.93%,111.94% | 61.61%, 82.52% | 61.58%, 82.49% | 61.58%, 82.49% | 61.58%, 82.49% |
| **InceptionResNetV2** | 5.83%, 9.14% | 4.1%, 7.2% | 3.58%, 6.47% | 3.58%, 6.47% | 3.57%, 6.48% | 3.57%, 6.49% |
| **InceptionV3** | 8.37%, 13.55% | 5.84%, 10.81% | 7.02%, 11.31% | 7.05%, 11.35% | 5.68%, 9.73% | 5.65%, 9.69% |
| **MobileNet** | 13.64%,23.64% | 13.26%, 23.34% | 13.38%, 23.57% | 11.12%, 20.83% | 11.12%, 20.83% | 11.12%, 20.83% |
| **MobileNetV2** | 23.27%, 34.15% | 21.45%, 32.29% | 21.75%, 32.88% | 21.39%, 32.37% | 21.39%, 32.37% | 21.39%, 32.37% |
| **NASNetLarge** | 21.34%, 23.57% | 18.2%, 20.42% | 18.52%, 20.87% | 18.54%, 20.9% | 18.49%, 20.83% | 18.54%, 20.9% |
| **NASNetMobile** | 32.93%, 35.29% | 19.15%, 21.44% | 19.75%, 22.44% | 19.79%, 22.49% | 19.83%, 22.54% | 20.01%, 22.76% |
| **ResNet50** | 17.72%, 31.5% | 17.05%, 30.59% | 16.79%, 31.51% | 17.32%, 30.82% | 17.32%, 30.82% | 17.32%, 30.82% |
| **ResNet152** | 9.76%, 22.84% | 9.52%, 20.54% | 9.5%, 20.81% | 9.49%, 20.83% | 9.49%, 20.83% | 9.49%, 20.83% |
| **xception** | 5.59%, 8.11% | 5.76%, 8.09% | 2.32%, 4.59% | 2.32%, 4.59% | 2.32%, 4.59% | 2.32%, 4.59% |
| **VGG16** | 49.67%, 76.84% | 28.76%, 49.02% | 32.31%, 49.74% | 32.31%, 49.74% | 32.31%, 49.74% | 32.31%, 49.74% |
| **VGG19** | 43.12%, 70.58% | 26.98%, 48.72% | 22.33%, 31.82% | 22.87%, 32.47% | 22.33%, 31.82% | 22.33%, 31.82% |
| **DenseNet201** | 25.03%, 51.8% | 24.16%, 48.68% | 24.52%, 48.89% | 24.52%, 48.88% | 24.52%, 48.89% | 24.52%, 48.89% |

Table 3. Distance between the points with minimum and maximum energy consumption in the whole search space (including 25 different point) in different array sizes

| PE-array / Network name | [12,14] $\Delta_{min}^{max}$ | [16,16] $\Delta_{min}^{max}$ | [32,32] $\Delta_{min}^{max}$ | [64,64] $\Delta_{min}^{max}$ | [128,128] $\Delta_{min}^{max}$ | [256,256] $\Delta_{min}^{max}$ |
|---|---|---|---|---|---|---|
| **AlexNet** | 65.12% | 58.76% | 60.44% | 60.44% | 60.44% | 60.44% |
| **DenseNet121** | 75.45% | 73.06% | 74.2% | 74.15% | 74.17% | 74.16% |
| **DenseNet169** | 91.07% | 88.86% | 89.71% | 89.69% | 89.72% | 89.72% |
| **GoogleNet** | 113.21% | 113.29% | 113.62% | 113.63% | 113.63% | 113.63% |
| **InceptionResNetV2** | 27.11% | 23.06% | 17.67% | 17.6% | 17.6% | 17.61% |
| **InceptionV3** | 31.14% | 25.96% | 20.7% | 20.79% | 18.64% | 18.57% |
| **MobileNet** | 32.06% | 25.06% | 26.11% | 24.64% | 24.81% | 24.81% |
| **MobileNetV2** | 30.4% | 29.81% | 30.31% | 29.98% | 30.04% | 30.04% |
| **NASNetLarge** | 38.05% | 33.38% | 33.89% | 33.91% | 33.83% | 33.91% |
| **NASNetMobile** | 46.82% | 30.21% | 30.56% | 30.61% | 30.64% | 30.85% |
| **ResNet50** | 48.17% | 34.02% | 34.29% | 34.32% | 34.43% | 34.45% |
| **ResNet152** | 33.11% | 32.04% | 32.52% | 32.55% | 32.55% | 32.55% |
| **xception** | 15.15% | 18.84% | 12.12% | 12.12% | 12.11% | 12.11% |
| **VGG16** | 92.62% | 62.31% | 51.65% | 51.65% | 51.65% | 51.65% |
| **VGG19** | 83.72% | 59.69% | 59.58% | 60.5% | 59.59% | 59.59% |
| **DenseNet201** | 62.87% | 61.04% | 61.46% | 61.45% | 61.47% | 61.47% |

Fig. 7 shows the accelerator's elapsed time of network processing of the VGG16 and ResNet50 in a fixed $GB_{ifmap}$ when sweeping $GB_{psum}$ for different array sizes. As stated before, the accelerator's total processing time and the energy consumption is a summation of the time and energy consumed in the memory hierarchy and the processing array. Obviously, by increasing array size and consequently increasing processing capacity, the array processing time is expected to be reduced.

Fig. *8* shows the processing time spent in the array for VGG16 with constant $GP_{psum}$ and $GB_{ifamp}$ when the array size increases.

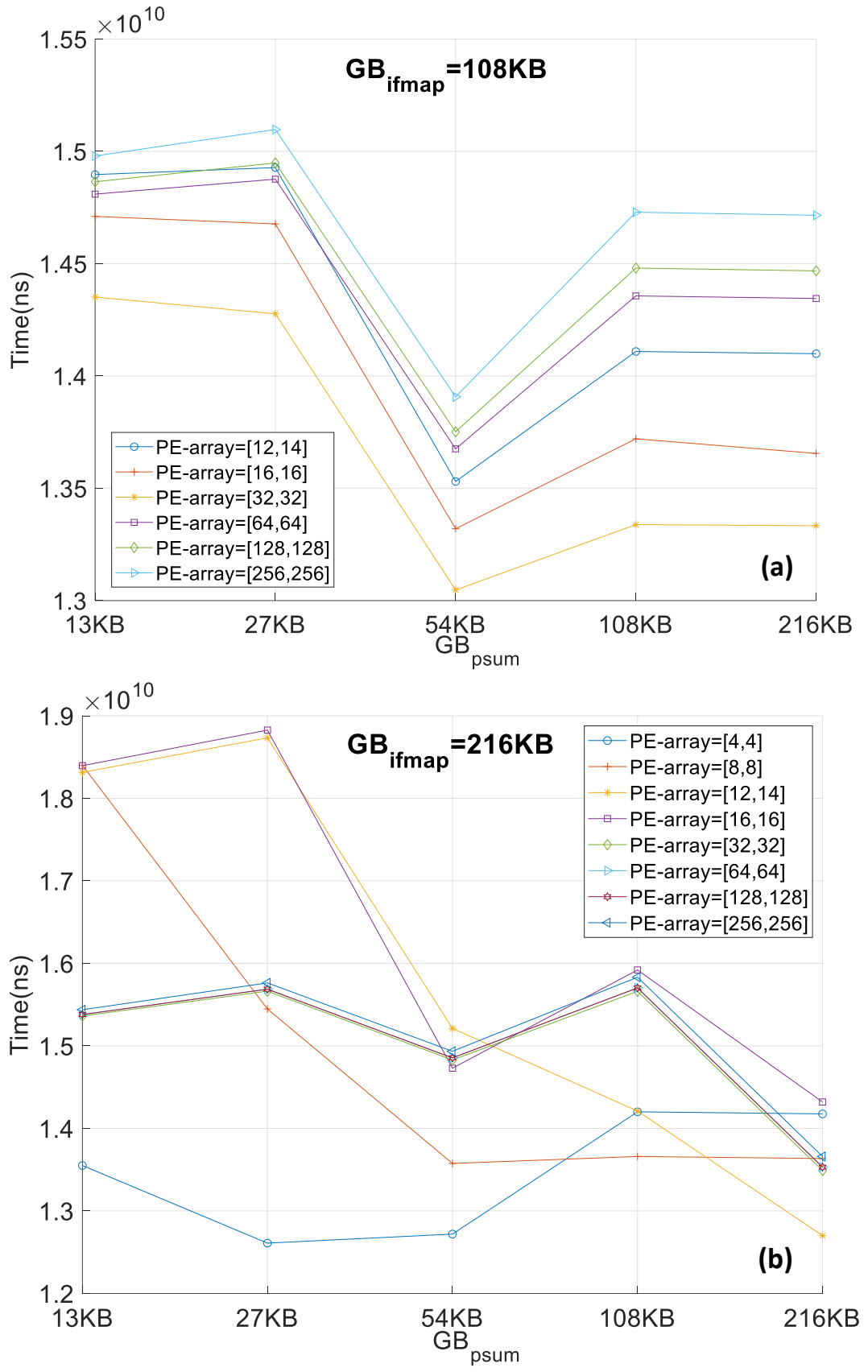

Fig. 7. Accelerator's elapsed time in a constant $GB_{ifmap}$ when sweeping $GB_{psum}$ for different array sizes for (a) ResNet50 (b) VGG16.

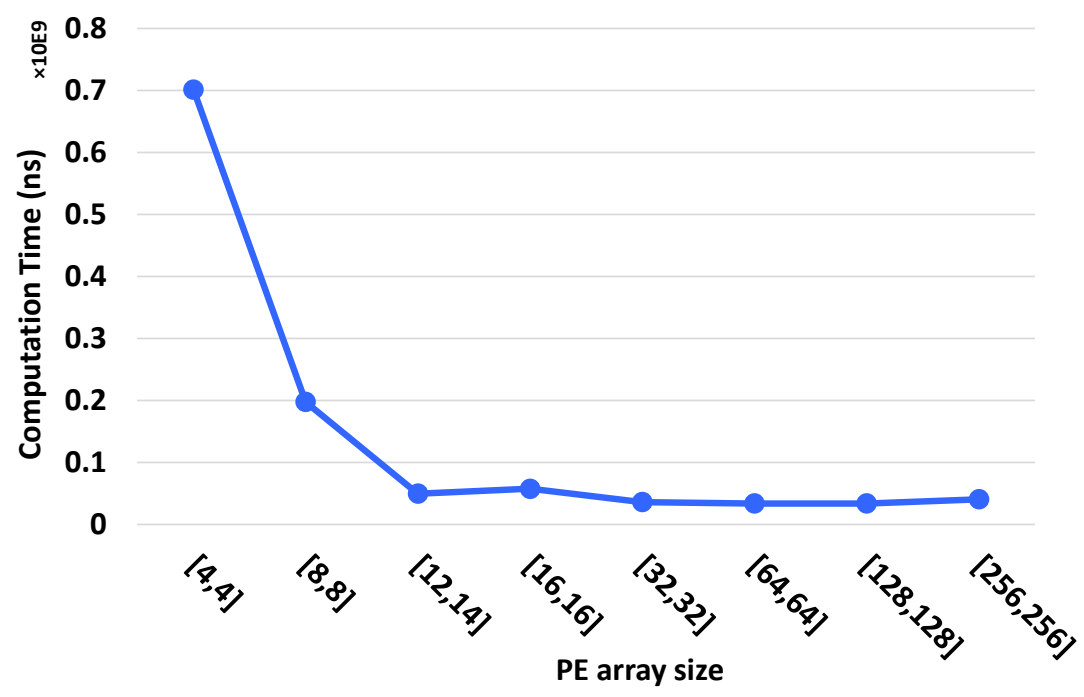

Fig. 8. The processing time spent in the array for VGG16 in a constant $GB_{ifmap}$ and a constant $GB_{psum}$ when the array size increases.

For example, when the array size increases from [4,4] to [8,8] and from [16,16] to [32,32], the processing time decreases by about 71.85% and 37.11%, respectively. In spite of the fact that the amount of time spent in the array decreases with increasing array size, the processing time behavior of the accelerator (by considering time spent in the memory hierarchy) is similar to the accelerator's overall energy consumption.This shows the importance and impact of the memory hierarchy in the architecture of accelerators designed for neural network processing.

***Observation 3:***

In a given specific dataflow, in a constant $GB_{ifamp}$ size, increasing the array size cannot reduce the processing elapsed time unless $GB_{psum}$ size is commensurate with the amount of partial sum produced by the array size. In other words, the array will only achieve its maximum performance if the data is provided with the least possible delay for processing.

Fig. *9* shows the accelerator's elapsed time in two constant $GB_{psum}$ when sweeping $GB_{ifmap}$ for different array sizes. In the case of low-capacity array sizes (e.g. [4,4]) increasing the $GB_{ifamp}$ size leads to an increase in elapsed time.

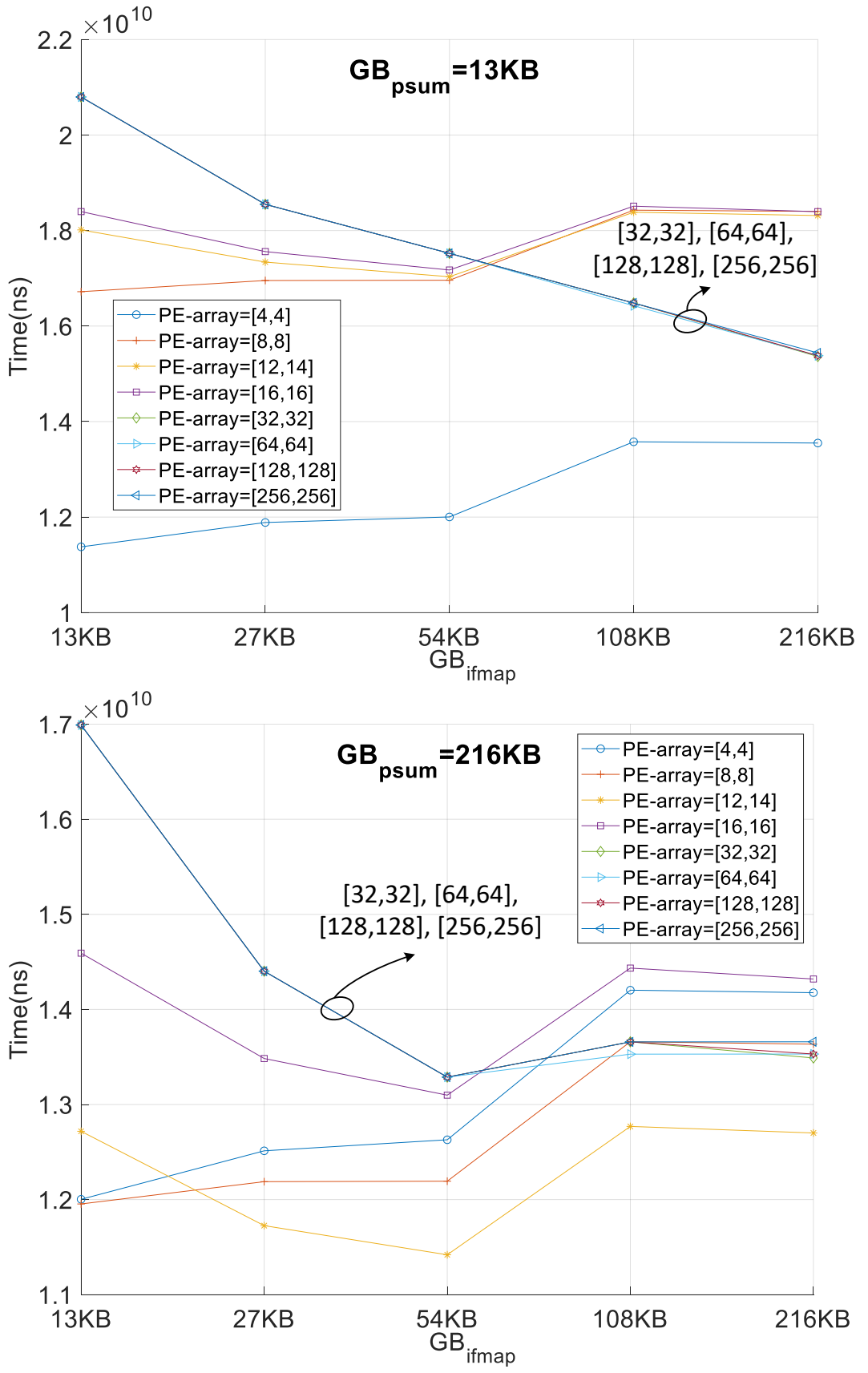

Fig. 9. Accelerator's elapsed time in two different constant $GB_{psum}$ when sweeping $GB_{ifmap}$ for different array sizes for VGG16.

The reason for this increase is the additional reading and writing time associated with larger buffers. In fact, due to the lower processing capacity (smaller array sizes), the amount of data required by the array for processing is low enough that even small $GB_{ifmap}$ can store it. In contrast, this trend is the opposite for larger arrays. This means as $GB_{ifmap}$ size increases, the elapsed time decreases. In fact, larger $GB_{ifmap}$ sizes can store the larger amount of data required by larger array sizes, which in turn leads to less write and re-read of partial sums to and from $GB_{psum}$ that could not otherwise be achieved with low-sized $GB_{ifmap}$.

***Observation 4:***

In a given specific dataflow, in order to reduce the processing time to the near-optimal point, the size of $GB_{ifmap}$ should be in proportion to the processing capacity of the array. A failure to choose the appropriate value will result in a high processing time cost and may cause the processing time to be longer than for smaller arrays.

## IV. HETEROGENEOUS MULTI-CORE CHIP FOR EFFICIENT PROCESSING OF NEURAL NETWORKS

In the previous section, we saw how changes in architectural elements can affect energy consumption and the processing time of neural accelerators. To have a general view of possible changes in the whole search space (including a total of 150 possible points), Table 4 shows two different parameters defined as follows for 18 different neural network topologies. First, the average difference between the EDP of each point of the search space and the EDP of the minimum point (the point in the whole search space where EDP is at its lowest value):

$$mean\left(\frac{EDP_i - EDP_{min}}{EDP_{min}}\right) \times 100, i = 1, 2, \ldots N \quad (4)$$

Second, the maximum of this difference in the whole search space is defined as follows:

$$max\left(\frac{EDP_i - EDP_{min}}{EDP_{min}}\right) \times 100, i = 1, 2, \ldots N \quad (5)$$

Where, N denotes the total number of points in the entire search space and $EDP_{min}$ denotes the minimum EDP value in the entire search space. The results in Table 4 show that even in a not-so-large search space moving away from the optimal point can be very costly in terms of both energy consumption and the performance of the accelerator. In other words, by choosing the proper architectural elements based on the neural network topology to be processed, it is possible to reduce energy consumption and increase the performance of the accelerator simultaneously. Hence, to illustrate this possibility, we define a scenario as follows. Suppose we are looking for a configuration in which the desired parameter of choice (for example EDP) is close to the minimum for each benchmark network in Table 4. In this regard, we define a 5% boundary around the minimum point. In other words, we first find the minimum point in the search space and consider all the other points in the search space whose EDP value is 5% higher than the minimum point. Table 5 shows all these configurations as well as the minimum point for different networks, where the minimum point is written in **bold**. Each column shows a configuration in which the EDP is in a 5% boundary.

Table 4. The mean difference between the EDP of each point and the minimum value of EDP in the whole search space alongside the maximum of this difference

| Network Name | $mean\left(\frac{EDP_i - EDP_{min}}{EDP_{min}}\right)$ (%) | $max\left(\frac{EDP_i - EDP_{min}}{EDP_{min}}\right)$ (%) |
|---|---|---|
| **AlexNet** | 78.7 | 257.6 |
| **DenseNet121** | 62.79 | 169.09 |
| **DenseNet169** | 57.58 | 219.77 |
| **DenseNet201** | 43.75 | 137.79 |
| **InceptionResNetV2** | 16.82 | 52.35 |
| **InceptionV3** | 21.52 | 61.01 |
| **ResNet50** | 36.13 | 104.24 |
| **ResNet50V2** | 40.81 | 142.57 |
| **ResNet101** | 30.87 | 68.23 |
| **ResNet152** | 29.57 | 73.77 |
| **VGG16** | 108.97 | 220 |
| **VGG19** | 97.43 | 223 |
| **GoogleNet** | 129.65 | 302.02 |
| **MobileNet** | 29.38 | 66.85 |
| **MobileNetV2** | 29.01 | 56.5 |
| **NASNetLarge** | 42.39 | 80.3 |
| **NASNetMobile** | 59.75 | 133.3 |
| **Xception** | 20.9 | 134.62 |

Table 5. All the configurations within 5% boundary of the minimum point. For a better understanding of the format used in this table please refer to the text

| Network name | EDP1 | EDP2 | EDP3 | EDP4 | EDP5 | EDP6 | EDP7 |
|---|---|---|---|---|---|---|---|
| **VGG16** | (216/27, {1}) | (**216/54**, {**1**}) | | | | | |
| **VGG19** | (216/27, {1}) | (**216/54**, {**1**}) | | | | | |
| **AlexNet** | (**54/54**, {3 **4** 5 6}) | | | | | | |
| **DenseNet121** | (54/13, {1 3 4 5 6}) | (**54/27**, {**1** 2 3 4 5 6}) | (54/54, {1 2 3 4 5 6}) | | | | |
| **DenseNet169** | (**54/27**, {**1** 2 3 4 5 6}) | (54/54, {1 2 3 4 5 6}) | | | | | |
| **DenseNet201** | (54/13, {1 3 4 5 6}) | (**54/27**, {**1** 2 3 4 5 6}) | (54/54, {1 2 3 4 5 6}) | | | | |
| **GoogleNet** | (216/27, {1 2}) | (**216/54**, {**1** 2 3 4 5 6}) | (216/108, {3 4 5 6}) | (216/216, {3 4 5 6}) | | | |
| **InceptionResNetV2** | (54/13, {1 2 3 4 5 6}) | (**54/27**, {1 **2** 3 4 5 6}) | (54/54, {1 2 3 4 5 6}) | (108/27, {1}) | (108/54, {1}) | (216/27, {1 2 3 4 5 6}) | (216/54, {1 3 4 5 6}) |
| **InceptionV3** | (54/13, {2 3}) | (54/27, {1}) | (54/54, {2 3 4 5 6}) | (108/27, {1}) | (108/54, {1}) | (**216/27**, {1 2 3 **4** }) | (216/54, {2 3 4 5 6}) |
| **MobileNet** | (54/13, {1 2 3 4 5 6}) | (54/27, {1 2 3 4 5 6}) | (54/54, {2}) | (216/27, {1 2 3}) | (**216/54**, {1 **2**}) | | |
| **MobileNetV2** | (54/13, {2 3 4 5 6}) | (54/27, {2 3 4 5 6}) | (216/13, {1 3 4 5 6}) | (**216/27**, {1 2 **3** 4 5 6}) | (216/54, {1 2 3 4 5 6}) | | |
| **NASNetLarge** | (216/27, {1 3 4 5 6}) | (**216/54**, {1 3 4 5 **6**}) | | | | | |
| **NASNetMobile** | (**216/54**, {3 4 5 **6**}) | | | | | | |
| **ResNet50** | (**54/13**, {**1** 2 3 4 5 6}) | (54/27, {1 2 3 4 5 6}) | (54/54, {1 2}) | | | | |
| **ResNet50V2** | (**54/27**, {**1**}) | (54/54, {1}) | | | | | |
| **ResNet101** | (**54/27**, {**1** 2 3 4 5 6}) | (54/54, {1}) | | | | | |
| **ResNet152** | (**54/27**, {**1** 2 4 5 6}) | (54/54, {1}) | | | | | |
| **xception** | (**27/54**, {3 4 5 6}) | (**216/54**, {**1** 2 3 4 5 6}) | | | | | |

Each configuration is in $(GB_{psum}/GB_{ifmap}\ ,\{index_{array}\})$ format. Both of the $GB_{ifmap}$ and $GB_{psum}$ are a member of {13KB, 27KB, 54KB, 108KB, 216KB} set and, $index_{array}$ refers to the index of the following set's members {[12,14], [16,16], [32,32], [64,64], [128,128], [256, 256]}. As an example, $index_{array}$ = 3 refers to the array of size [32,32]. In cases where more than one $index_{array}$ is written in a column, it means all of those array sizes are within 5% bound. By looking at the different possible configurations in the table, you can find commonalities between them. Here, we consider two different configurations $(54/54,\{3\})$ and $(216/54,\{1\})$ for processing two sets of networks. The first category includes AlexNet, DenseNet family, and ResNet family which will be processed on the core with $(54/54,\{3\})$ configuration. The second category includes VGG family, MobileNet family, NASNet and xception family that will be processed on the core with the $(216/54,\{1\})$ configuration. As the two selected configurations are among the near-optimal configurations of the InceptionResNetV2 and InceptionV3, each can be processed with either core. By processing the networks in a configuration that is close to their optimal configuration, energy loss will be largely avoided.

Table *6* shows the increase in energy consumption (denoted by $\Delta_{\boldsymbol{E}}$), processing time (denoted by $\Delta_{\boldsymbol{D}}$) , and EDP (denoted by $\Delta_{\boldsymbol{EDP}}$) of the benchmarked networks when they are processed on their non-corresponding cores (cores that are selected to process another group of networks). As an example, the EDP value is about 63.72% higher when the AlexNet is processed with the core with $(216/54,\{1\})$ configuration rather than $(54/54,\{3\})$ configuration. Or in another example, we see that processing GoogleNet on the core with $(54/54,\{3\})$ configuration leads to 67.69% higher EDP compared to the case when a core with $(216/54,\{1\})$ is used. On average, we will observe 16.45% and 30% higher EDP values when the first category networks are processed on the core configuration specified for the second category and vice versa, respectively.

Table 6. The increase in energy consumption, processing time, and EDP of networks when they are processed on a non-corresponding core

| Network Name | $\Delta_E$ (%) | $\Delta_D$ (%) | $\Delta_{EDP}$ (%) |
|---|---|---|---|
| VGG16 | 33.33 | 25.97 | 67.96 |
| VGG19 | 26.26 | 18.34 | 49.42 |
| GoogLeNet | 36.82 | 22.75 | 67.96 |
| MobileNet | 2.2 | 3.49 | 5.77 |
| MobileNetV2 | 2.94 | 1.4 | 4.38 |
| NASNetLarge | 14.51 | 8.51 | 23.87 |
| NASNetMobile | 11.42 | 3.44 | 15.25 |
| xception | 4.11 | 1.8 | 6 |
| **Network Name** | **$\Delta_E$ (%)** | **$\Delta_D$ (%)** | **$\Delta_{EDP}$ (%)** |
| AlexNet | 26.78 | 29.13 | 63.72 |
| DenseNet121 | 8.21 | 7.41 | 16.23 |
| DenseNet169 | 8.19 | 8.52 | 17.42 |
| DenseNet201 | 6.86 | 6.91 | 14.25 |
| ResNet50 | 4.83 | 2.96 | 7.94 |
| ResNet50V2 | 2.79 | 1.8 | 4.7 |
| ResNet101 | 3.72 | 0.25 | 4 |
| ResNet152 | 3.5 | 0.13 | 3.36 |

*A. Heterogeneous scheme for neural network accelerator*

In the previous section, we discussed how improper selection of values in a specific search space of the processing core's elements, can result in inefficiencies in both the accelerator's energy consumption and the performance. Therefore, the "one processing core fits all" solution is not optimal in terms of energy consumption/performance efficiency. Accordingly, we recommend a multi-core processor with heterogeneous cores for near-optimal processing. Fig. 10 shows an overview of the proposed accelerator. In this example, the processor uses two different types of cores, each with its own array size and memory hierarchy (represented in the figure with different patterns). Each core reads the required data from the external main memory or writes the processed data to it through its own global buffer. Having more than one identical processing core of each core type (homogeneous cores) allows us to use model parallelism in order to increase the processing speed.

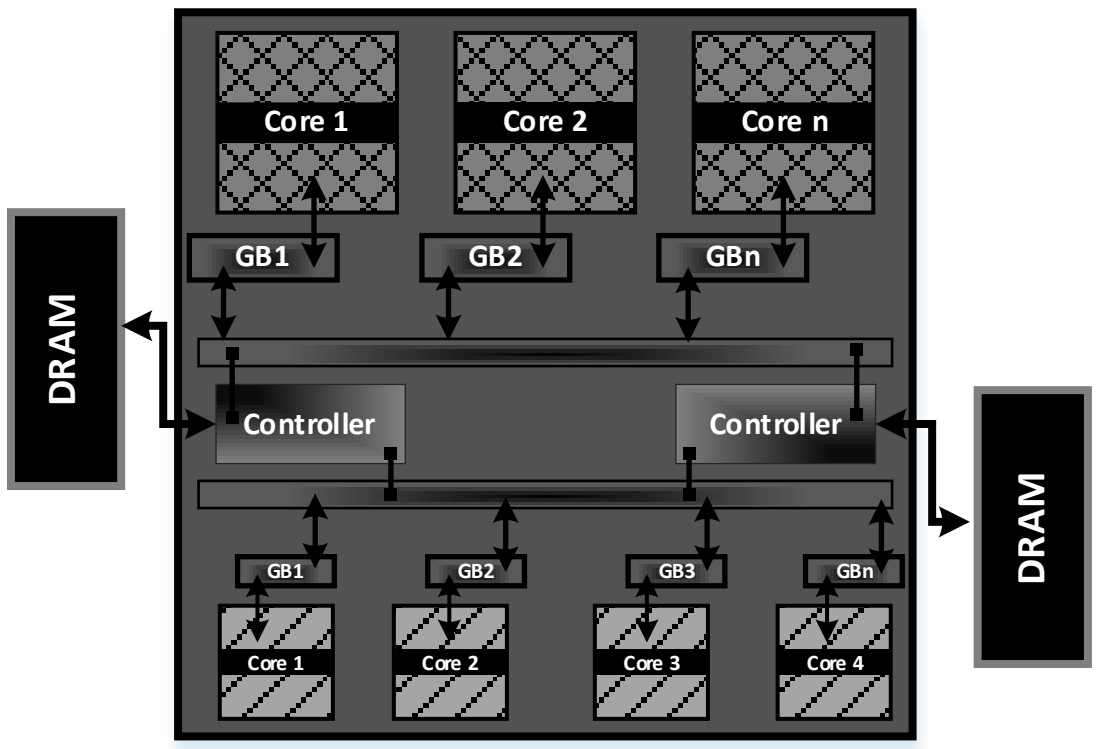

Fig. 10. Proposed accelerator for neural network processing with two groups of cores.

In order to determine the number of heterogeneous core groups, we must first calculate the target parameter that we intend to minimize during network processing by the accelerator. For example, if we are attempting to minimize the energy consumption of the accelerator, the energy consumption of processing the networks is extracted for each point in the search space. It should be noted that the search space consists of all possible combinations of architectural elements. In the next step, by defining a definite boundary from the point where the desired parameter is minimal, we consider all possible configurations within the boundary as candidates for the core which is responsible for processing the network. This procedure increases the probability of finding common configurations between different networks. In exploring all of these configurations for all networks, we select common configurations on which the maximum number of networks can be processed. This study leads to two or a few more different configurations for processing cores.

*B. model parallelism on homogeneous cores*

Neural networks are processed layer by layer. To increase the processing speed, the network's layers can be distributed across a certain number of cores and processed in parallel. In this pipeline of processing cores, each writes its processed data to the off-chip main memory and the next core reads its input data

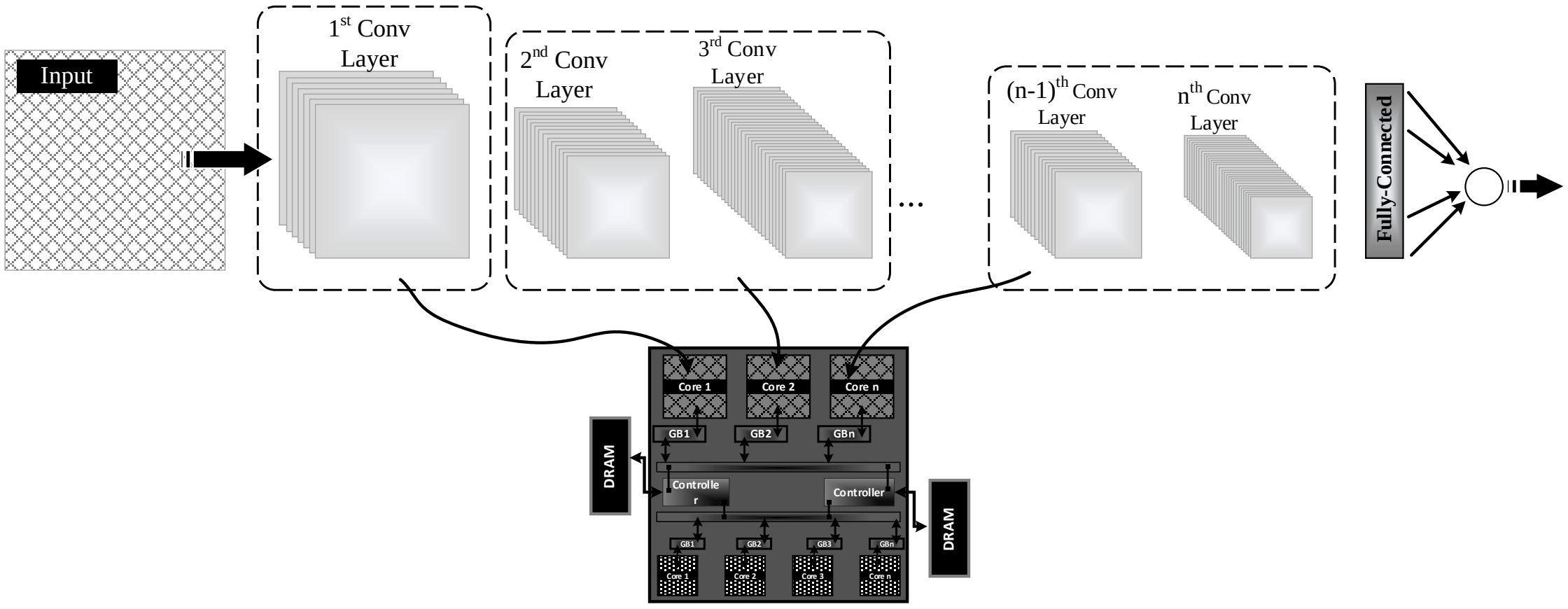

Fig. 11. Distribution of a network across a number of homogeneous cores.

**ALGORITHM II** BRANCH-AND-BOUND USED FOR MODEL PARALLELISM

$n = 1, best_pipeline_{latency} = inf, average = \frac{\Sigma D_{array}}{\#Cores}, d_{sum} = 0$

branch_and_bound($D_{array}$, $average$, $n$):

  While($d_{sum} < \mu_d$):

    $d_{sum} = d_{sum} + D_{array}[i]$

    i++

  End

If ($d_{sum} > best_pipeline_{latency}$) //acting as the bound

  Return

If($! = end\ of\ layers$)

  $core_{latency} = sum(D_{array}[0{:}\,i])$

//assigning $layer_0$ $to\ layer_i$ $to\ the\ core$

from it. Fig. 11 illustrates a distributed network on a few cores. For maximum efficiency, the processing load of each of these cores should be as close as possible to one another. Evidently, imbalanced load distribution would result in lower utilization and performance. Therefore, to achieve load balancing between the cores, we utilize a branch-and-bound algorithm whose pseudocode is illustrated in Algorithm II.

As shown in the pseudocode, the algorithm requires two inputs. One is the processing latency of the layers (estimated by the Tool), and the other is the number of cores specified by the user. These two inputs are used to calculate the average latency of each core in the optimal state (i.e. the result of dividing the total latency of the layers by the number of cores) which will be used as a measure of the balanced distribution of processing load throughout the algorithm.

The algorithm starts from the first layer and progresses to the layer in which the total latency of these layers exceeds the average (the algorithm stops at the first layer that meets this condition). Afterward, two branches are formed, one including the last layer and the other not. Evidently, in the first branch, the total latency of the layers is greater than the optimal average (measurement criterion), and for the second branch, it is less than that. In the next step, for each branch formed in the previous step, the first and second steps are repeated for the remaining layers in the array until the final layer of the network is reached. It should be noted that after a round of calculating the latency for each core (reaching the end of the layers and assigning each layer to the respective cores), the maximum latency is considered to be the latency of the cores (in pipeline architecture, the largest latency is considered as the latency of the entire pipeline). It is worth noting that when a new branch is created, if a core's latency exceeds this value, the calculations for that branch will be halted ("bound" condition). In fact, this branch with a latency greater than the so-far calculated latency cannot be considered to be the optimal value. Therefore, further calculations will not be necessary.

To verify the correct operation of the proposed algorithm, we use the two configurations selected in Table 5. We assume that there are **three** processing cores with $GB_{psum}$ = 54KB, $GB_{ifmap}$ = 54KB, and array size of [32,32] and **four** processing cores with $GB_{psum}$ = 216KB, $GB_{ifmap}$ = 54KB, and array size of [12,14] in the chip. Table 7 and Table 8 show how the first and second categories of networks are distributed across three and four cores, respectively. The first number in the tuple $(l_{initial}, n_C)$ indicates the layer number of the network that will be assigned to the corresponding processing core as the first layer to be processed, whereas the second number indicates the number of layers to which the core will be assigned for processing. The speedup in the tables is calculated as follows:

$$Speedup = \frac{latency\ of\ single\ core\ excecution}{\max(latency\ of\ each\ core\ in\ multicore\ excecution)} \quad (6)$$

Table 7. Distribution of network layers across three homogeneous processing cores with near-optimal configuration alongside the speedup of this parallelism

| **Network Name** | **Core #1** $(l_{initial}, n_C)$ | **Core #2** $(l_{initial}, n_C)$ | **Core #3** $(l_{initial}, n_C)$ | **S**** |
|---|---|---|---|---|
| **AlexNet** | (1, 1) | (2, 1) | (3, 3) | 2.01 |
| **DesneNet121** | (1, 11) | (12, 24) | (36, 85) | 2.81 |
| **DesnseNet169** | (1, 14) | (15,7 0) | (85, 84) | 2.72 |
| **DenseNet201** | (1, 33) | (34, 82) | (116, 85) | 2.98 |
| **InceptionResNetV2** | (1, 54) | (55, 46) | (101, 144) | 2.94 |
| **InceptionV3** | (1, 28) | (29, 31) | (60, 35) | 2.92 |
| **ResNet50** | (1, 18) | (19, 18) | (37, 17) | 2.94 |
| **ResNet50V2** | (1, 19) | (20, 18) | (38, 16) | 2.93 |
| **ResNet101** | (1, 36) | (37, 34) | (71, 34) | 2.95 |
| **ResNet152** | (1, 54) | (55, 51) | (106, 50) | 2.96 |

**Stands for speedup

Table 8. Distribution of network layers across four homogeneous processing cores with near-optimal configuration alongside the speedup of this parallelism

| Network Name | Core #1 $(l_{initial}, n_C)$ | Core #2 $(l_{initial}, n_C)$ | Core #3 $(l_{initial}, n_C)$ | Core #4 $(l_{initial}, n_C)$ | S** |
|---|---|---|---|---|---|
| **VGG16** | (1, 3) | (4, 3) | (7, 9) | (10, 4) | 3.12 |
| **VGG19** | (1, 4) | (5, 3) | (8, 3) | (11, 6) | 3.72 |
| **GoogleNet** | (1, 11) | (12, 11) | (23, 17) | (40, 18) | 3.74 |
| **MobileNet** | (1, 13) | (14, 11) | (25, 8) | (33, 9) | 3.81 |
| **MobileNetV2** | (1, 12) | (13, 27) | (40, 16) | (56, 14) | 3.88 |
| **NASNetLarge** | (1, 6) | (7, 330) | (337, 46) | (383, 106) | 3.89 |
| **NASNetMobile** | (1, 1) | (2, 26) | (28, 232) | (260, 27) | 3.73 |
| **xception** | (1, 4) | (5, 1) | (6,1) | (7,68) | 2.34 |
| **InceptionResNetV2** | (1, 33) | (34, 50) | (84, 38) | (122, 123) | 3.9 |
| **InceptionV3** | (1, 18) | (19, 24) | (43, 25) | (68, 27) | 3.92 |

**stands for speedup

## V. Conclusion

In this work, we studied the impact of architectural elements changes on the performance and energy efficiency of array-based accelerators using an in-house developed tool. Our observations indicated that the impact of different elements is highly intertwined with each other. In other words, the design of an efficient accelerator requires a microscopic understanding of each element's impact on the final energy consumption and processing latency of the accelerator as it executes neural networks. In view of these interactions, we proposed a multi-core processing chip scheme for the efficient execution of a wide variety of deep neural networks. The evaluations within a selective search space demonstrated that up to 67% of energy-delay product could be saved if the neural networks were run on their near-optimal core configurations. Additionally, we suggested a simple branch-and-bound algorithm for distributing the network across the cores. By employing this algorithm on different benchmarks in designed scenarios, near-optimal speedup values were achieved.